\documentclass[journal=jacsat,manuscript=article]{achemso}

\usepackage[version=3]{mhchem} 
\usepackage{xr}
\usepackage[colorlinks=true,urlcolor=blue,linkcolor=blue,citecolor=blue]{hyperref}

\usepackage{lineno}
\modulolinenumbers[5]
\usepackage{chemformula} 
\usepackage{natbib}
\usepackage{graphicx}
\usepackage{float}
\usepackage{caption}
\usepackage{booktabs}
\usepackage{soul}
\usepackage{amssymb}
\usepackage{subfigure}
\usepackage[marginal]{footmisc}
\usepackage{filecontents}

\author{Jing Ning}
\affiliation[SUTD]
{Singapore University of Technology and Design (SUTD), 8 Somapah Road, 487372, Singapore}
\alsoaffiliation[NUS]
{Department of Materials Science \& Engineering, National University of Singapore, 9 Engineering Drive 1, 117575, Singapore}
\altaffiliation{Contributed equally to this work}
\email{jing_ning@mymail.sutd.edu.sg}
\author{Yunzheng Wang}
\affiliation[SUTD]
{Singapore University of Technology and Design (SUTD), 8 Somapah Road, 487372, Singapore}
\altaffiliation{Contributed equally to this work}
\author{Ting Yu Teo}
\affiliation[SUTD]
{Singapore University of Technology and Design (SUTD), 8 Somapah Road, 487372, Singapore}

\author{Chung‑Che Huang}
\affiliation[UK]
{Optoelectronics Research Centre, University of Southampton, Southampton SO17 1BJ, UK}
\author{Ioannis Zeimpekis}
\affiliation[UK]
{Optoelectronics Research Centre, University of Southampton, Southampton SO17 1BJ, UK}
\author{Katrina Morgan}
\affiliation[UK]
{Optoelectronics Research Centre, University of Southampton, Southampton SO17 1BJ, UK}

\author{Siew Lang Teo}
\affiliation[IMRE]
{Institute of Materials Research and Engineering (IMRE), Agency for Science Technology and Research (A*STAR), 2 Fusionopolis Way, Innovis, Singapore}

\author{Daniel W. Hewak}
\affiliation[UK]
{Optoelectronics Research Centre, University of Southampton, Southampton SO17 1BJ, UK}

\author{Michel Bosman}
\affiliation[NUS]
{Department of Materials Science \& Engineering, National University of Singapore, 9 Engineering Drive 1, 117575, Singapore}
\alsoaffiliation[IMRE]
{Institute of Materials Research and Engineering (IMRE), Agency for Science Technology and Research (A*STAR), 2 Fusionopolis Way, Innovis, Singapore}

\author{Robert E. Simpson}
\affiliation{Singapore University of Technology and Design (SUTD), 8 Somapah Road, 487372, Singapore}
\email{robert_simpson@sutd.edu.sg}
\title
  {Low energy switching of phase change materials using a 2D thermal boundary layer}

\abbreviations{IR,NMR,UV}
\keywords{Phase change memory, Van der Waals interfaces, 2D materials, Thermal engineering}

\begin{document}

\begin{abstract}
The switchable optical and electrical properties of phase change materials (PCMs) are finding new applications beyond data storage in reconfigurable photonic devices. 
However, high power heat pulses are needed to melt-quench the material from crystalline to amorphous.
This is especially true in silicon photonics, where the high thermal conductivity of the waveguide material makes heating the PCM energy inefficient.
Here, we improve the energy efficiency of the laser induced phase transitions by inserting a layer of two dimensional (2D) material, either \ch{MoS2} or \ch{WS2}, between the silica or silicon and the PCM. 
The 2D material reduces the required laser power by at least 40\% during the amorphization (RESET) process, depending on the substrate.
Thermal simulations confirm that both \ch{MoS2} and \ch{WS2} 2D layers act as a thermal barrier, which efficiently confines energy within the PCM layer.
Remarkably, the thermal insulation effect of the 2D layer is equivalent to a $\sim$100~nm layer of \ch{SiO2}.
The high thermal boundary resistance induced by the van der Waals (vdW)-bonded layers limits the thermal diffusion through the layer interfaces.
Hence, 2D materials with stable vdW interfaces can be used to improve the thermal efficiency of PCM-tuned Si photonics devices.
Furthermore, our waveguide simulations show that the 2D layer does not affect the propagating mode in the Si waveguide, thus this simple additional thin film produces a substantial energy efficiency improvement without degrading the optical performance of the waveguide.
Our findings  pave the way for energy-efficient laser-induced structural phase transitions in PCM-based reconfigurable photonic devices.
\end{abstract}

\pagebreak
\section{Introduction}
Phase change materials (PCMs) have been commercialized for both optical and electrical data storage because they exhibit a large optical and electrical property contrast that can be induced in nanoseconds, and once switched, the properties are latched into a metastable state; i.e. they exhibit non-volatility\citep{wuttig2007phase,lencer2008map,siegrist2011disorder,kolobov2004understanding}. 
The data RESET process is achieved by converting the crystalline phase to the amorphous phases using heat by either a short current or a laser pulse to melt the PCM.
The molten state is then quenched at a high rate to freeze in the disordered state\citep{xiong2011low,zallo2016laser}.  
The switching is reversible and can be repeated billions of times\citep{zhang2019designing}. 
The reversible transition from the amorphous to the crystalline phase (SET) is induced by heating the material to a temperature above the glass transition temperature and below the melting temperature for a relatively longer time.

PCMs have already been used in commercial products, from DVD-RW optical discs\citep{ohta2001phase} to advanced 3D X'Point electrical memory\citep{joelhruska}, and now they are widely studied for universal memory and neuro-inspired computing\citep{tuma2016stochastic,zhang2019designing}.
Despite the commercial successes and potential photonic applications, the amorphization operation can be energy inefficient because the heat easily dissipates into the surroundings.
Typically, only $\sim$1\% of the supplied energy is used by the phase change material, and this energy inefficiency limits the potential applications of phase change materials\citep{sadeghipour2006phase}.

Many efforts have been devoted to improving the energy efficiency in chalcogenide PCM switching. 
From an electrical device perspective, one effective way is to reduce the contact area between the bottom electrode and the phase change material cell to decrease the switching volume of phase change materials.
This is done by replacing the typical mushroom structure with edge-contact-type\citep{ha2003edge}, bridge-type\citep{chen2006ultra}, or $\mu$Trench\citep{pellizzer200690nm} structures, or by applying a nanoscale electrode such as carbon nanotubes\citep{xiong2011low}.
Material optimizations can also improve the device switching energy efficiency.
Doping is a good approach to modify the properties of Ge-Sb-Te ternary phase change materials, for example, by using Sc\citep{wang2018scandium}, Ti\citep{zhu2014uniform}, C\citep{zhou2014understanding}, Cr\citep{wang2015cr} etc. 
These dopant atoms diffuse into the phase change materials or  partially substitute atoms to form local defects or distortions, which prevent the nucleated crystals from growing into large grains. 
Strong phonon scattering from the additional grain boundaries results in low thermal conductivity, which enables generated heat in the PCM to be trapped within it. 
This decreases the energy transfer to the surroundings and low-energy switching can be realized. 
However, incorporating dopants often produces phase separation, which is seen as the material is cyclically switched between its amorphous and crystalline states, thus shortening the lifespan of phase change memory cells\citep{xia2015ti,borisenko2011understanding}.
Stacking two different PCMs in a superlattice-like structure can also lower the thermal conductivity and resultant switching energy\citep{chong2006phase}. 
The lower thermal conductivity is achieved at the interfaces, which are produced by the alternating layers within the layered structure.
Interfacial phase change materials (iPCMs) have also been used to lower entropic losses during the phase transition\citep{simpson2011interfacial}.
Strain engineering these iPCMs, by exploiting the lattice mismatch of the \ch{Sb2Te3} and \ch{GeTe} layers, was applied to further enhance the energy performance of the iPCM\citep{kalikka2016strain,zhou2017avalanche}. 
However, growing the iPCM  requires accurate control of the physical vapor deposition system conditions.
A more straightforward approach to lowering the programming energy of electrical phase change memory devices involves inserting an interfacial layer between phase change materials and heater electrodes. 
Indeed, \ch{Ta2O5}\citep{matsui2006ta2o5}, fullerene\citep{kim2008fullerene}, \ch{WO3}\citep{rao2008programming}, and \ch{TiO2}\citep{xu2008lower} have all shown some promise at decreasing the programming voltages in electrical devices. 
The low power switching originates from the low thermal conductivity of these inserted layers but the downside of this approach is that the inserted layer increases the electrical resistance of the electrical memory, which negatively affects the overall device performance. 

Most of the efforts to increase the switching energy performance of PCMs have focused on electrical memory devices. 
However, with the increasing interest in phase change material programmable photonics, we need to start considering how to make these devices switch efficiently too.
Some of the most studied applications of PCM-programmable photonics include Si waveguides or plasmonic metamaterials\citep{rude2013optical, rios2015integrated, wuttig2017phase, cao2013rapid,lu2021reversible}.
In both cases the PCM is typically interfaced directly with high thermal conductivity materials, both of which have a large thermal conducitivty. 
For example, the thermal conductivities of silicon and gold are respectively 140~W$\cdot$m$^{-1}\cdot$K$^{-1}$ and 318~W$\cdot$m$^{-1}\cdot$K$^{-1}$.
Two dimensional (2D) van der Waals (vdWs) materials, such as graphene\citep{ahn2015energy} and \ch{MoS2}\citep{neumann2019engineering}, are known to have a low out-of-plane thermal conductivity and are, therefore, interesting to study as a way to increase the thermal boundary resistance (TBR) between a material with high thermal conductivity, such as a silicon waveguide, and a PCM.
Indeed, others have shown that the RESET energy of PCM electrical memories can be lower when a 2D material is placed at the interface of the electrical heater and the PCM, however the enhancement is shadowed by high in-plane thermal transport, thus additional fabrication steps are needed to limit the contact area\citep{ahn2015energy,neumann2019engineering}.
Until now, the effect of 2D materials on the switching energy performance of PCM-based photonic devices has not been studied but we hypothesize that incorporating 2D materials into photonic devices will efficiently trap the heat in the PCM and could radically lower the switching energy.
Moreover, since these 2D materials are optically thin, we do not expect them to influence the optical performance of the device. 

In PCM-integrated photonic devices, the PCM provides a means to route and attenuate light in a photonic circuit\citep{rude2013optical, rios2015integrated, wuttig2017phase}.
For PCM-integrated photonics devices, refractive index switching is realized by applying heat pulses to the PCM, which is in contact with the waveguide.
However, higher power laser pulses are needed to introduce the structural phase transitions and concomitant refractive index changes in the PCM when it is directly interfaced with a Si waveguide due to its high thermal conductivity. 
When these PCM switches are incorporated in a large-scale photonic network with an array of interconnected waveguide meshes, the energy needed to precisely program the network will be high and scale unfavorably with the number of PCM-tuned elements and this will ultimately limit the network scalability.
Indeed, in hardware neural networks, a single programming pulse energy should be in the fJ range\citep{wang2020resistive}, but current PCM programming pulse energy on Si waveguides is in the pJ range\citep{rude2013optical}. 
Thus, for these devices to become practical, we must start considering how to make them thermally energy efficient.
Since silica-on-silicon and Si substrates are often respectively used in plasmonics and photonic integrated circuits, we study how 2D layers of \ch{MoS2} and \ch{WS2} on silica-on-silicon and Si substrates influence the laser energy required to switch a PCM. 


In this work, the 2D material was placed as an atomically thin interfacial thermal barrier underneath the PCM, either on a silicon substrate, or on a silica-on-silicon substrate, as shown in Figure~\ref{fig_1}(a) and (b).
Since the inert vdWs interfaces do not have any dangling bonds, we expect they do not affect the structural transition behavior of phase change materials. 
We expect that a few atomic layers ($\sim$1-2 nm) of a dielectric material will not change the optical performance of photonic devices. 
Thermally, on the other hand, their effect is expected to be sizable, as the weak vdWs interfaces of the transition metal dichalcogenides (TMDCs) 2D layer should strongly limit heat transport along the out-of-plane direction\citep{kim2018disorder}.
Hence, energy will be confined within the small volume of phase change material and greatly reduce the power used to switch the PCM. 
We study by experiment and simulation whether this inherent thermal property is common to two different TMDCs 2D materials, \ch{MoS2} and \ch{WS2}, and show that these 2D TMDC layers are indeed effective at increasing the optical switching power efficiency in PCM-tuned Si photonic devices.
We believe that this design is applicable to a wide range of PCM-based photonic devices, including thin-film reflective displays\citep{dong2019wide}, programmable plasmonic devices\citep{cao2013rapid}, and metasurfaces\citep{lu2021reversible}.

\section{Methods}
\subsection{Growth and characterization}
The stacked sandwich structure consists of either a 300$\pm$15~nm silica-on-silicon or a non-oxidized Si substrate, a 2D TMDC thermal barrier, and a GeTe phase change material layer, as shown in Figure~\ref{fig_1}(a). 
The \ch{MoS2} and \ch{WS2} were prepared using Atomic Layer Deposition (ALD) for the silica-on-silicon substrate and Van der Waals Epitaxy (VdWE) for the Si substrate.
For the \ch{MoS2} growth, the substrate was treated in a UV/\ch{O3} reactor for 10 minutes. 
After that, \ch{MoO3} was grown using thermal ALD in a Cambridge Nanotech Savannah S200 system using bis(tert-butylimido)-bis(dimethylamido) molybdenum as a molybdenum precursor at 250 $^{\circ}$C. 
The films were then sulfurized in a tube furnace using an \ch{H2S}/Ar gas mixture with a final annealing temperature of 970~$^{\circ}$C. 
The in-house developed VdWE apparatus was used to grow monolayer \ch{WS2}\citep{felix2020comprehensive,gordo2018revealing}. 
\ch{WCl6} (99.9\% pure from Sigma-Aldrich) was used as the precursor, kept in a bubbler, delivered by Ar gas to VdWE system to react with \ch{H2S} gas to form \ch{WS2} monolayer on the substrates at the set growth temperatures of 900~$^{\circ}$C.
A deposition time of 5 minutes was required to achieve uniform \ch{WS2} monolayer films.
Subsequently, \ch{GeTe} was deposited from a \ch{GeTe} alloy target (2'' diameter and 99.999\% pure from AJA International) using magnetron sputtering (AJA Orion5) in an Ar atmosphere with a pressure of 3.7 mtorr (0.5 Pa) with a working distance of 140 mm. 
The deposition rate was 0.95 nm/min using a power of 10~W for 1880~s.
We also placed the blank substrates along with the \ch{MoS2}$/$\ch{WS2} deposited samples to act as the control samples.
The sample structure was measured using Transmission Electron Microscopy (TEM, FEI Titan) with an acceleration voltage of 200~kV, as shown in Figure~\ref{fig_1}(b). 
Focused Ion Beam (FIB, FEI Helios Nanolab 450S) milling was necessary to prepare the lamella for cross-sectional TEM image.
The crystalline GeTe films were first prepared by annealing the as-deposited GeTe film using a temperature-controlled heating stage at 300~$^{\circ}$C for 10 min.
The crystallization temperature was found by differentiating the reflectivity curve, which was recorded whilst heating the samples from room temperature to 300 $^{\circ}$C with a 4 $^{\circ}$C/min ramp rate (Linkam Scientific Instruments Ltd). 
To protect the film from oxidation, which is known to influence its phase transitions\citep{zhou2016oxygen}, the anneal was performed in an Ar atmosphere flowing at 4 SCCM.
Raman spectra were collected at room temperature using a WITec Alpha300R system equipped with a 633-nm wavelength excitation; the incident laser intensity was kept low to minimize irradiation-induced heating of the probed region. 
The thickness measurement was carried out via Atomic Force Microscopy (AFM, Asylum Research, MFP-3D Origin).
Our in-house developed static tester, which consists of a low-power 638-nm probe laser and a relatively high-power 660-nm pump laser, was used to measure the switching power and time\citep{behera2017laser}. 
The system can simultaneously measure the reflection of the probe laser from the sample whilst the pump laser pulses heat the sample. 
The focused laser spot had a beam size of 0.8~$\mu$m ($1/{e^2}$ intensity) on the sample. 
Here, we used the static tester to laser write an array of crystallization and amorphization marks under different laser pulse widths and incident powers. 
The reflected signal from the probe laser was collected before and after the pump pulses.

\subsection{Finite difference simulations}
The heat induced by laser pulses can increase the temperature of PCMs and achieve crystallization or amorphization. 
The transient temperature profile is obtained from the unsteady heat conduction equation, as given in Equation~\ref{eq_1},
\begin{equation}
	\rho c\frac{\partial{T(x,y,z,t)}}{\partial t}= \triangledown\cdot\kappa\triangledown T(x,y,z,t)+Q(x,y,z,t)
	\label{eq_1}
\end{equation}
where, $T(x, y, z, t)$ is the temperature at a site of $(x, y, z)$ and a certain time $t$, $\rho$ is the mass density, $c$ is the specific heat capacity, $\kappa$ is the thermal conductivity, $Q(x, y, z, t)$ is the Joule heat brought by the laser pulse, which can be expressed as Equation~\ref{eq_2}, assuming a Gaussian beam profile,
\begin{equation}
	Q(x,y,z,t)=e^{-\alpha z}\frac{2P_{in}}{\pi\omega^2}(1-R)\alpha e^{-2\frac{x^2+y^2}{\omega^2}}f(t)
	\label{eq_2}
\end{equation}
where, $P_{in}$ is the laser power, $\omega$ is the $1/e^2$ Gaussian beam radius, $\alpha$ is the absorption coefficient, $R$ is the reflectivity, and $f(t)$ is the temporal waveform. 

Here, $\rho$ of GeTe ($\sim$6.19~g$\cdot$cm$^{-3}$) and $c$ of GeTe ($\sim$259.2~J$\cdot$kg$^{-1}\cdot$K$^{-1}$) were used\citep{LandoltBornstein1998:sm_lbs_978-3-540-31360-1_834}.
Meanwhile, $\rho$ of \ch{MoS2} ($\sim$5.06~g$\cdot$cm$^{-3}$), $c$ of \ch{MoS2} ($\sim$379.6~J$\cdot$kg$^{-1}\cdot$K$^{-1}$)\citep{haynes2015crc,suryavanshi2019thermal}, $\rho$ of \ch{WS2} ($\sim$7.5~g$\cdot$cm$^{-3}$) and $c$ of \ch{WS2} ($\sim$250~J$\cdot$kg$^{-1}\cdot$K$^{-1}$) were used\citep{haynes2015crc,ruppert2017role}, respectively.
$R$ of a-GeTe (0.44), c-GeTe (0.68), a-GeTe/\ch{MoS2} (0.47), c-GeTe/\ch{MoS2} (0.64), a-GeTe/\ch{WS2} (0.54), c-GeTe/\ch{WS2} (0.66) were measured with a 660-nm laser.
$\alpha$ of a-GeTe ($\sim$1.8$\times$10$^7~$m$^{-1}$) and c-GeTe ($\sim$4.75$\times$10$^7$~m$^{-1}$) were calculated from their extinction coefficient\citep{jafari2016zero}.
We used $\kappa$ of a-GeTe ($\sim$0.204 W$\cdot$m$^{-1}\cdot$K$^{-1}$) and c-GeTe ($\sim$3.59~W$\cdot$m$^{-1}\cdot$K$^{-1}$)\citep{ghosh2020thermal}.
Thermal boundary conductance of \ch{MoS2} and \ch{WS2} layers were used $\sim$16~MW$\cdot$m$^{-2}\cdot$K$^{-1}$ and $\sim$5.5~MW$\cdot$m$^{-2}\cdot$K$^{-1}$\citep{yu2020plane,gertych2021phonon}.
The thickness of the GeTe was $\sim$30~nm and $\omega$ was measured as 0.8~$\mu$m. 
100-ns and 500-ns laser pulses with different power were applied in amorphization and crystallization simulations, respectively.

\subsection{Waveguide Simulation}
The PCM-tuned Si waveguides used in the optical simulation were optimized in the transverse electric (TE) mode. 
The dimensions were chosen to ensure single mode operation\citep{teo2022comparison,xu2019low}. 
In the simulation model, refractive indices of the Si waveguide, \ch{MoS2}/\ch{WS2} and \ch{GeTe} layers were obtained from literature\citep{EDWARDS1985,liu2020temperature,wuttig2017phase}. 
The refractive index values can be found in the supporting information Figure~\ref{fig_S5}. 
To obtain the mode profile and overall effective index values of the waveguide, we solved Maxwell’s equations on the waveguide cross section using the Finite Difference Eigenmode solver from Lumerical Mode Solution (LMS).

\section{Results and Discussion}

\begin{figure}
\centering
\includegraphics[width=\textwidth]{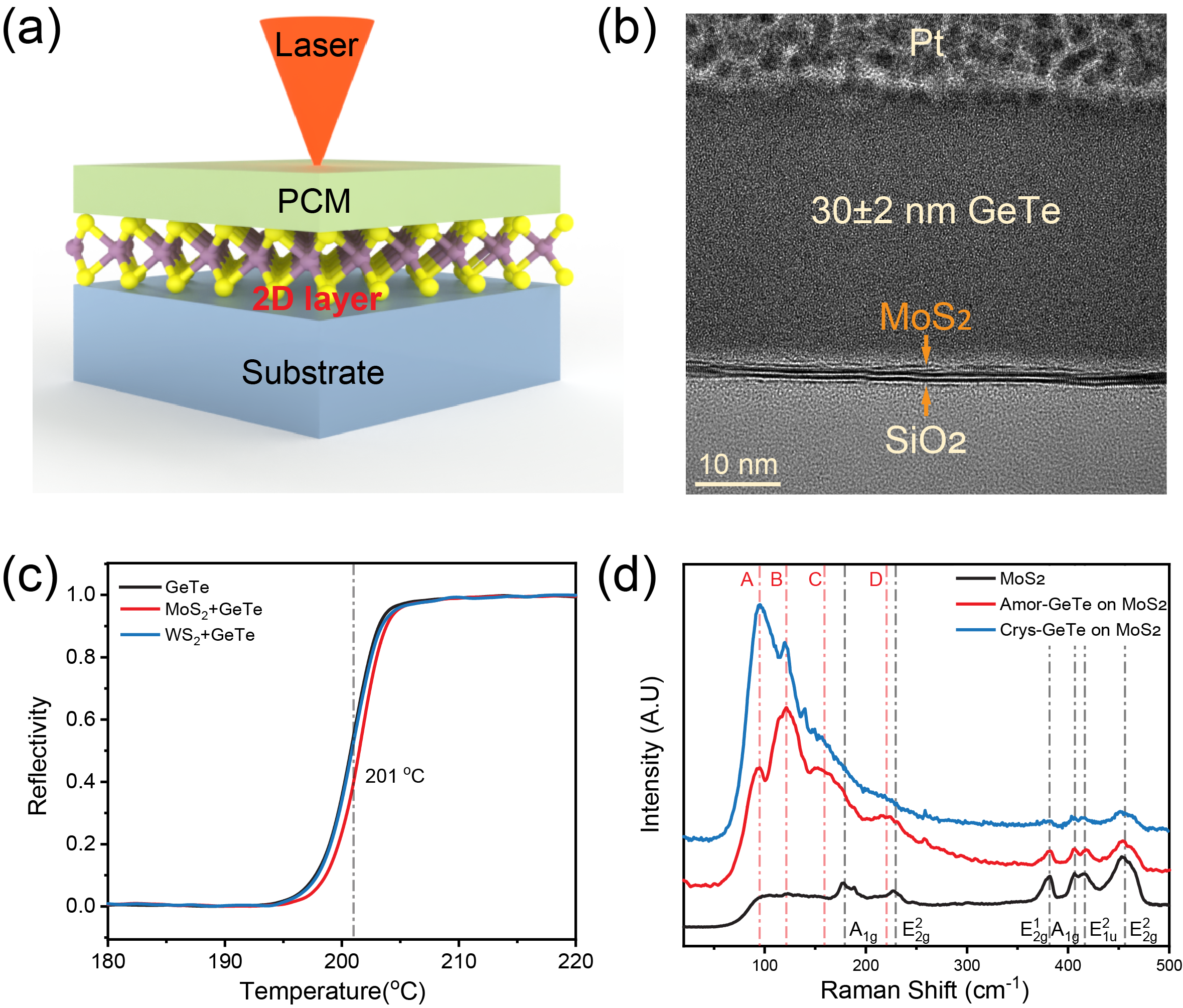}
\caption{(a) Schematic of the programmable optical sandwich consisting of substrate, 2D layer, and phase change material.
(b) Cross-sectional TEM image of interfacial bilayer \ch{MoS2} between the GeTe film and the silica-on-silicon substrate.
(c) Crystallization temperature of only GeTe, GeTe on \ch{MoS2}, or GeTe on \ch{WS2}, all on the silica-on-silicon substrate.
(d) Raman spectra of only \ch{MoS2}, as-deposited amorphous GeTe on \ch{MoS2}, and annealed crystalline GeTe on \ch{MoS2}, all on the silica-on-silicon substrate. 
The phonon modes for \ch{GeTe} and \ch{MoS2} are highlighted and labeled using red and black, respectively.
}
\label{fig_1}
\end{figure}

To ascertain that the 2D material layer is chemically inert and does not influence the structural transformation of GeTe, we measured the crystallization temperature and phonon modes of the GeTe on top of the TMDC 2D layers. 
The crystallization temperature of amorphous GeTe films, which were deposited directly on top of the crystalline bilayer \ch{MoS2}, monolayer \ch{WS2}, and the silicon substrate, was measured by recording the reflected intensity of visible light from the films as a function of temperature, as shown in Figure~\ref{fig_1}(c).
The sudden increase in reflectivity corresponds to the crystallization of the material.
Both the as-deposited GeTe sample and the GeTe on \ch{MoS2} or \ch{WS2} samples crystallized at 201$^{\circ}$C.
The consistent crystallization temperature indicates that the 2D materials do not influence the GeTe phase transition. 
This is expected since the vdW interfaces of the 2D layers are chemically inert and stable. 
Hence, the GeTe layer is physically isolated from the 2D material layer as no dangling bonds are present to form strong covalent bonds with the subsequent GeTe layer. 

To further confirm that the \ch{MoS2} layer does not affect the local structural transformation in GeTe crystallization, we also performed a Raman analysis. 
The Raman spectra of bilayer-\ch{MoS2} before and after GeTe deposition, and as-deposited GeTe film with \ch{MoS2} layers after annealing are presented in Figure~\ref{fig_1}(d).
In Figure~\ref{fig_1}(d), we highlight the GeTe and \ch{MoS2} phonon modes in red and grey dot-and-dash lines, respectively. 
The $A_{1g}$(179 cm$^{-1}$), $E^2_{2g}$(230 cm$^{-1}$), $E^1_{2g}$(382 cm$^{-1}$), $A_{1g}$ (408 cm$^{-1}$), $E^2_{1u}$(417 cm$^{-1}$), and $E^2_{2g}$(456 cm$^{-1}$) modes seen in the \ch{MoS2} sample were reported in the literature\citep{article,windom2011raman}.
The GeTe on \ch{MoS2} sample spectrum consists of a combination of amorphous GeTe and \ch{MoS2} peaks.
The as-deposited amorphous GeTe peaks occur at A (92 cm$^{-1}$), B (123 cm$^{-1}$), C (162 cm$^{-1}$), D (218 cm$^{-1}$) in the frequency range of 50-250 cm$^{-1}$\citep{andrikopoulos2006raman,zhou2016oxygen}. 
Upon crystallization, we observe a weaker signal in bands C (162 cm$^{-1}$) and D (218 cm$^{-1}$).
This indicates a local structural change of Ge from a lower tetrahedral coordination to an octahedral coordination, thus confirming that GeTe crystallization has occurred.
The \ch{MoS2} peaks are weakened by the 30-nm GeTe layer absorbing a portion of the scattered intensity. 
This effect is more substantial in the crystalline GeTe sample due to its higher absorption coefficient, which is induced by a denser and more compact crystalline structure upon annealing. 
From the spectral measurements, we see that the \ch{MoS2} Raman modes are unaffected by the GeTe layer and are able to withstand the deposition and heat-induced crystallization process. 
Hence, we conclude that the \ch{MoS2} layers can be used in conjunction with telluride PCMs without any further alternation of the layers bonds or stoichiometry. 
Moreover, the thermal stability of \ch{MoS2} with Te-based PCMs also indicates that other optimization strategies, such as superlattice or strain engineering, may be used for further switching energy efficiency improvements.
  
\begin{figure}
\centering
\includegraphics[width=\textwidth]{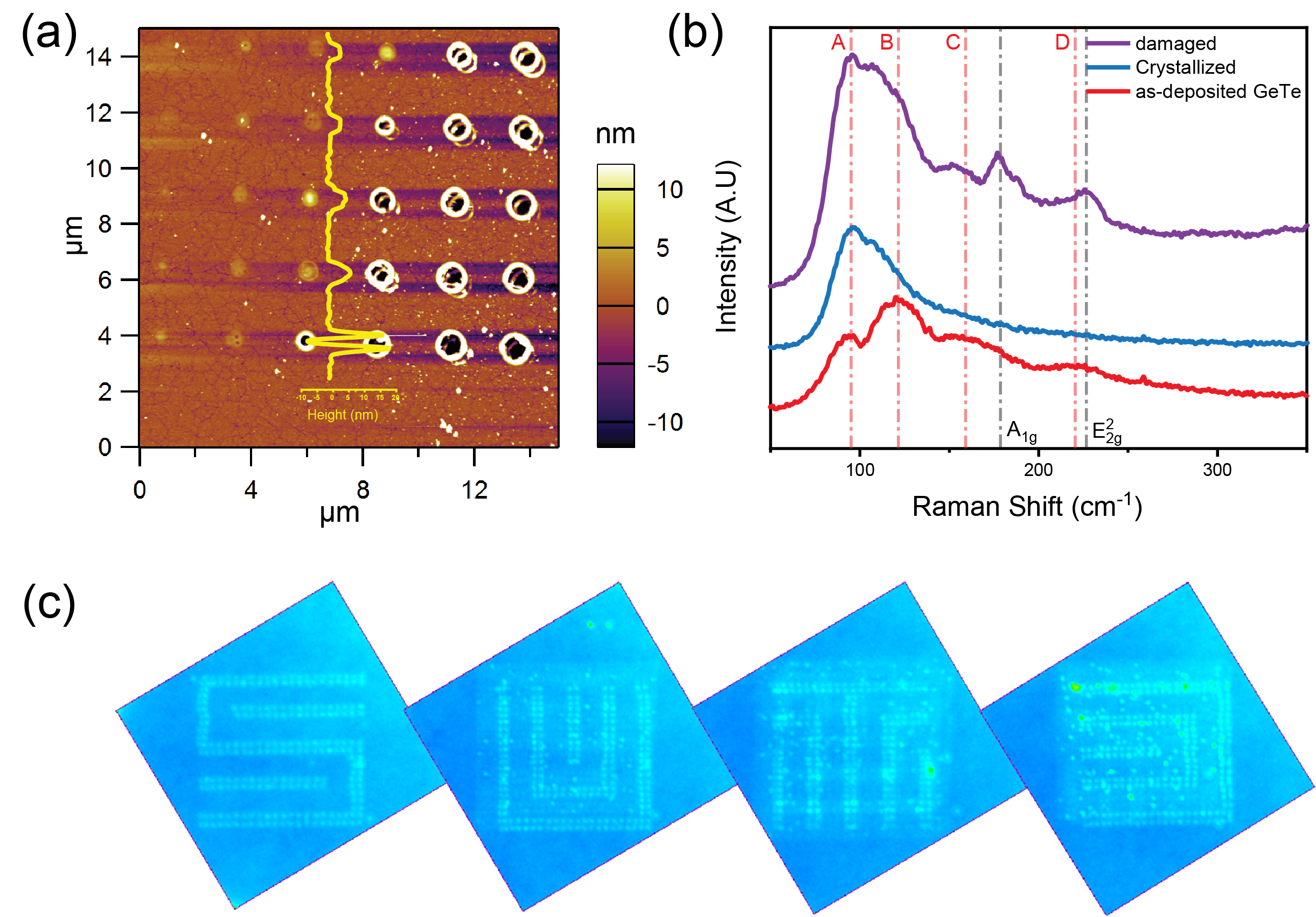}
\caption{Switching behavior of GeTe film on the silica-on-silicon substrate with \ch{MoS2} thermal barrier. 
(a) AFM topography of the write-mark matrix written into GeTe on the \ch{MoS2} bilayer on the silica-on-silicon substrate.
(b) Raman spectra of the GeTe films on the \ch{MoS2} bilayer in different structural states.
Red, blue and purple curves correspond to as-deposited amorphous, optically crystallized and ablated region.
The signals of GeTe and \ch{MoS2} are labeled using red and black color in the range of 50--350 cm$^{-1}$. 
(c) Optical images of re-amorphized laser-amorphized "S", "U", "T" and "D" characters, which were sequentially written into the same area of a recrystallized GeTe film. 
}
\label{fig_2}
\end{figure}

The aim of this work is to study whether the TMDC 2D layers can lower the heat power required for GeTe layers on thermally conductive substrates to crystallize or amorphize. 
To test this, we used a laser to write amorphous marks into the crystalline GeTe film on \ch{MoS2} and \ch{WS2} layers.
An amorphous-mark pulse power-time-reflectivity matrix was made by controlling the laser power and pulse duration. 
A microscope image of the resultant amorphization matrix for crystalline GeTe on \ch{MoS2} is shown in Figure~\ref{fig_S1}(b).
The laser pulse power was set in the range of 0--33 mW and the pulse duration was from 10 to 100 ns.
A visible reflectivity change was observed in the optical microscope image when the laser power reached 13.45~mW. 
The white and blue areas are where the GeTe amorphized or ablated, respectively.
We used an AFM to confirm whether the material truly amorphized or whether it had ablated, see Figure~\ref{fig_2}(a).
We observe a $\sim$2~nm increase in thickness upon amorphization, which corresponds to a $\sim$7\% thickness change between crystalline and amorphous states; 
a result that is consistent with previous reports\citep{njoroge2002density}.
GeTe ablation occurred for pulse powers above 17.6 mW and pulse durations longer than 70 ns. 
Ablation is visible as a slackening and squeezing from the center of the irradiated mark, which results in a micro-basin forming, as shown in the inset line profile on the AFM height map.
We then compare the Raman spectra of GeTe surfaces that consist of as-deposited regions, optically crystallized regions, and ablated regions, which were damaged by high power laser pulses.
The Raman spectra are shown in Figure~\ref{fig_2}(b).
The signals from the laser crystallized GeTe regions match the annealed crystalline GeTe sample, as shown in Figure~\ref{fig_1}(d).
We observe that the ablated regions have a stronger \ch{MoS2} signal because the GeTe surface has been removed and it is unable to efficiently absorb the \ch{MoS2} Raman scattered photons.

To demonstrate that GeTe on \ch{MoS2} can be reversibly switched and that consistently low power laser pulses can be used for amorphization, we amorphized dot-matrix-images of the characters "S", "U", "T", "D" from a crystalline region of the sample using low power laser pulses. 
Figure~\ref{fig_2}(c) shows optical micrographs of our rewritable pattern where the annealed crystalline GeTe on \ch{MoS2} sample was selected as the rewritable canvas. 
The blue background shows crystalline GeTe while the lighter shade of blue indicates GeTe amorphization.
To make this pattern, we used a 50 ns, 15.5 mW pulses to amorphize the film, and a 800-ns pulse with 7.46 mW pulse to crystallize it. 
To obtain a more observable contrast, these laser pulse powers are 15\% and 35\% higher than the minimum power required to amorphize and crystallize the GeTe sample, compromising the lifespan of GeTe to a certain extent.

\begin{figure}
\centering
\includegraphics[width=\textwidth]{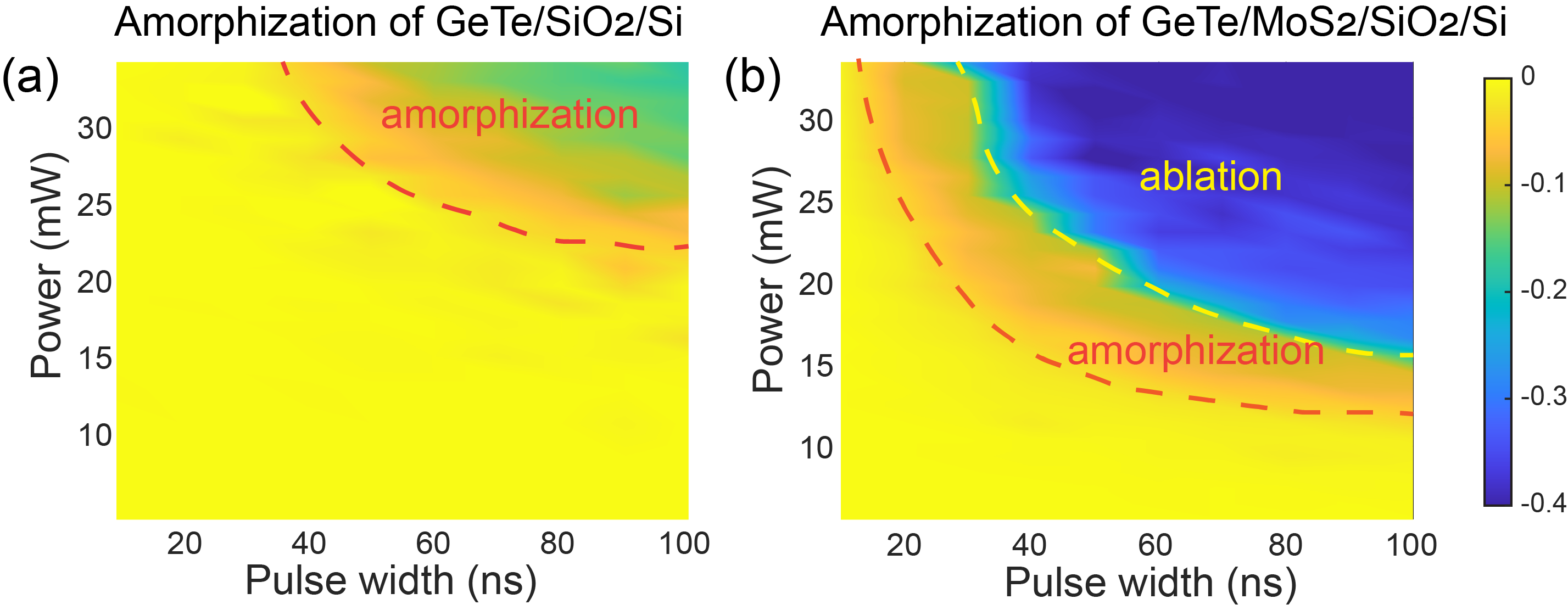}
\caption{Laser amorphization power-time-reflectivity measurement of (a) only GeTe and (b) GeTe with \ch{MoS2} on the silica-on-silicon substrate.
}
\label{fig_3}
\end{figure}

Thus far we have shown that interfacing \ch{MoS2} with GeTe does not influence structural transformations and we have confirmed that the GeTe can be amorphized and recrystallized by laser switching.
We now quantify the enhancement caused by the \ch{MoS2} 2D layer on the switching energy of GeTe films using our laser static tester\citep{behera2017laser}.
The laser static tester was used to amorphize the crystalline GeTe film with different pulse powers and lengths. 
The optical contrast gradually appeared with increasing pulse power and width (Figure~\ref{fig_S1}), indicated by a decrease in reflectivity, as shown in Figure~\ref{fig_3}(a) and (b).
For a GeTe layer deposited on a silica-on-silicon substrate, the amorphization power threshold was 21.95 mW for 70 ns pulses.
In contrast, amorphizing the GeTe interfaced with the \ch{MoS2} 2D layer on a silica-on-silicon substrate only required 13.45 mW and 50 ns, as shown in Figure~\ref{fig_3}(b).
The laser switching power used to amorphize the samples was reduced by 40\% by adding an \ch{MoS2} 2D layer. 
The greatly reduced amorphization power is attributed to the ultra-high thermal boundary resistance of the \ch{MoS2} interfaces, which confines the laser heat inside the small volume of the GeTe such that it rapidly reaches its melting temperature.
Moreover, if a 33.42~mW pulse is used to amrophize the GeTe on \ch{MoS2}, then the amorphization time is reduced by 67\% from 30 ns to 10 ns. 
Similarly, GeTe on \ch{WS2} on top of a silica-on-silicon substrate also produced a decrease in switching energy and time.
We direct the interested reader to Figure~\ref{fig_S2}, where the corresponding power-time-reflectivity plots for laser amorphization are included. 

It is interesting that the switching energy performance enhancement due to the TMDC layer is only seen for amorphization and not for crystallization.
We found that the GeTe with and without \ch{MoS2} is crystallized by a pulse with the same power.
This effect is due to the lower laser pulse power and longer time required for crystallization.
This means heat can diffuse further through the stacked layers into the substrate, which results in a smaller temperature gradient through the sample. 
Since the thermal conductivity of \ch{SiO2} is relatively low at 1.4~W$\cdot$m$^{-1}\cdot$K$^{-1}$, this limits the heat loss to some extent and makes the switching energy reduction unapparent.

\begin{figure}
\centering
\includegraphics[width=\textwidth]{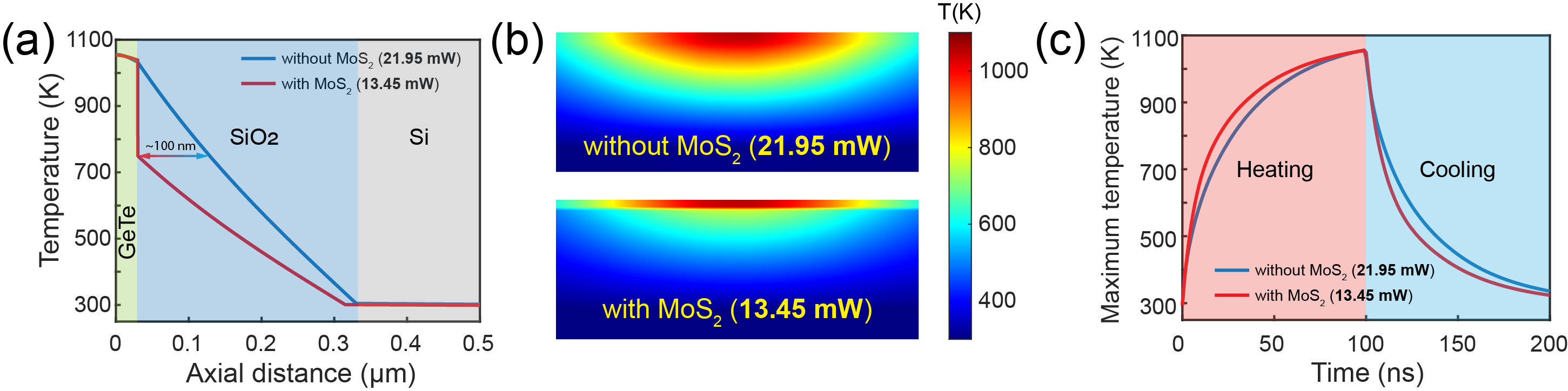}
\caption{Heat transport simulation of GeTe on \ch{MoS2} on a silica-on-silicon substrate during laser amorphization. 
(a) Axial temperature distribution in GeTe films on a silica-on-silicon substrate without and with \ch{MoS2} interfacial layers using different laser powers.
(b) Temperature distribution in cross-section of GeTe and GeTe on \ch{MoS2} samples with different power pulses.
(c) Maximum temperature on the surface of GeTe and GeTe on \ch{MoS2} samples after a 100 ns laser pulse with different powers.
}
\label{fig_4}
\end{figure}

We have found that adding 2D TMDC layers between GeTe and a substate is effective at lowering the RESET (amorphisation) power of GeTe without influencing its local atomic structure nor the crystallization temperature. 
We hypothesize, therefore, that the TMDC 2D layers must introduce an enormous TBR, and it is this TBR that improves the switching energy efficiency by trapping heat in the PCM. 
To study how \ch{MoS2} and \ch{WS2} layers can act as an efficient thermal boundary, we performed finite-difference simulations to model the heat transport between the  interfaces of the substrate/TMDC/PCM stack.
We simplified the model to a sandwich structure consisting of substrate, an infinitely thin 2D TBR interfacial layer, and the GeTe phase change material. 
The heat transport simulation for the amorphization and crystallizaiton process are presented in Figure~\ref{fig_4} and Figure~\ref{fig_S3}.
We modeled GeTe amorphization using the measured threshold power for amorphization with 100 ns pulses.
We see from the amorphization matrices in Figure~\ref{fig_S1} that the threshold laser powers are 13.45 mW and 21.95 mW with and without the \ch{MoS2} layer respectively.
Both the GeTe and the GeTe with \ch{MoS2} samples reached a similar temperature after 100 ns as shown in Figure~\ref{fig_4}, and this temperature is above the GeTe melting temperature, which is necessary for amorphization.
Importantly, the modeled laser power necessary to melt GeTe on \ch{MoS2} is 40\% less than that rquired to melt GeTe on an silica-on-silicon substrate. 
The temperature distribution plots in Figure~\ref{fig_4}(b) show that the \ch{MoS2} layer causes the heat  to be efficiently confined within the PCM layer. 
Moreover, the GeTe with \ch{MoS2} experiences a higher heating rate during the 100-ns laser pulse, and higher quench rate after the pulse ended, as shown in Figure~\ref{fig_4}(c).
At first glance this high quench rate may seem counterintuitive because the \ch{MoS2} interfacial layer has a large TBR.
However, the substrate temperature is much lower when the \ch{MoS2} layer is included, see Figure~\ref{fig_4}(b), and there is less thermal energy provided to the whole structure.
This means that only the GeTe layer needs to cool substantially, and the \ch{SiO2} can act as a heat sink and absorb the small amount of thermal energy that is trapped in the GeTe layer.
We conclude that just 1~nm of \ch{MoS2} can effectively prevent heat transfer to the substrate during 100 ns laser pulses. 
Indeed, in terms of thermal isolation, the 1~nm thick \ch{MoS2} is equavalent to $\sim$100 nm of \ch{SiO2}.

In the previous analysis, we simulated the heating using the different  laser pulse powers for amorphizaton.
However, if the same laser power is used for samples with and without the 2D TMDC layers, and if the pulse times are longer we can see that the samples reach thermal equilibrium at different temperatures. 
For example, we simulated a 5.54 mW 500 ns laser pulse in Figure~\ref{fig_S3}. 
This pulse condition is similar to that required for GeTe crystallization.
\ch{MoS2} causes the GeTe to equilibrate at 700 K. 
In constrast, the \ch{MoS2} on a silica-on-silicon substrate saturates at 600 K.
Both temperatures are above the 573~K required for GeTe crystallization but since GeTe crystallization is limited by the nucleation time, we do not see a significant difference in the overall crystallization time in the experiments.
The sample with a $\sim$300~nm thick \ch{SiO2} layer and a \ch{MoS2} layer equilibrates at 700~K rather than 600~K, which occurs without the \ch{MoS2} layer. 
However, we would expect the difference in equilibrating temperatures to be much more pronounced for TMDCs interfaced directly with a highly thermally conductive Si waveguide.

\begin{figure}
\centering
\includegraphics[width=\textwidth]{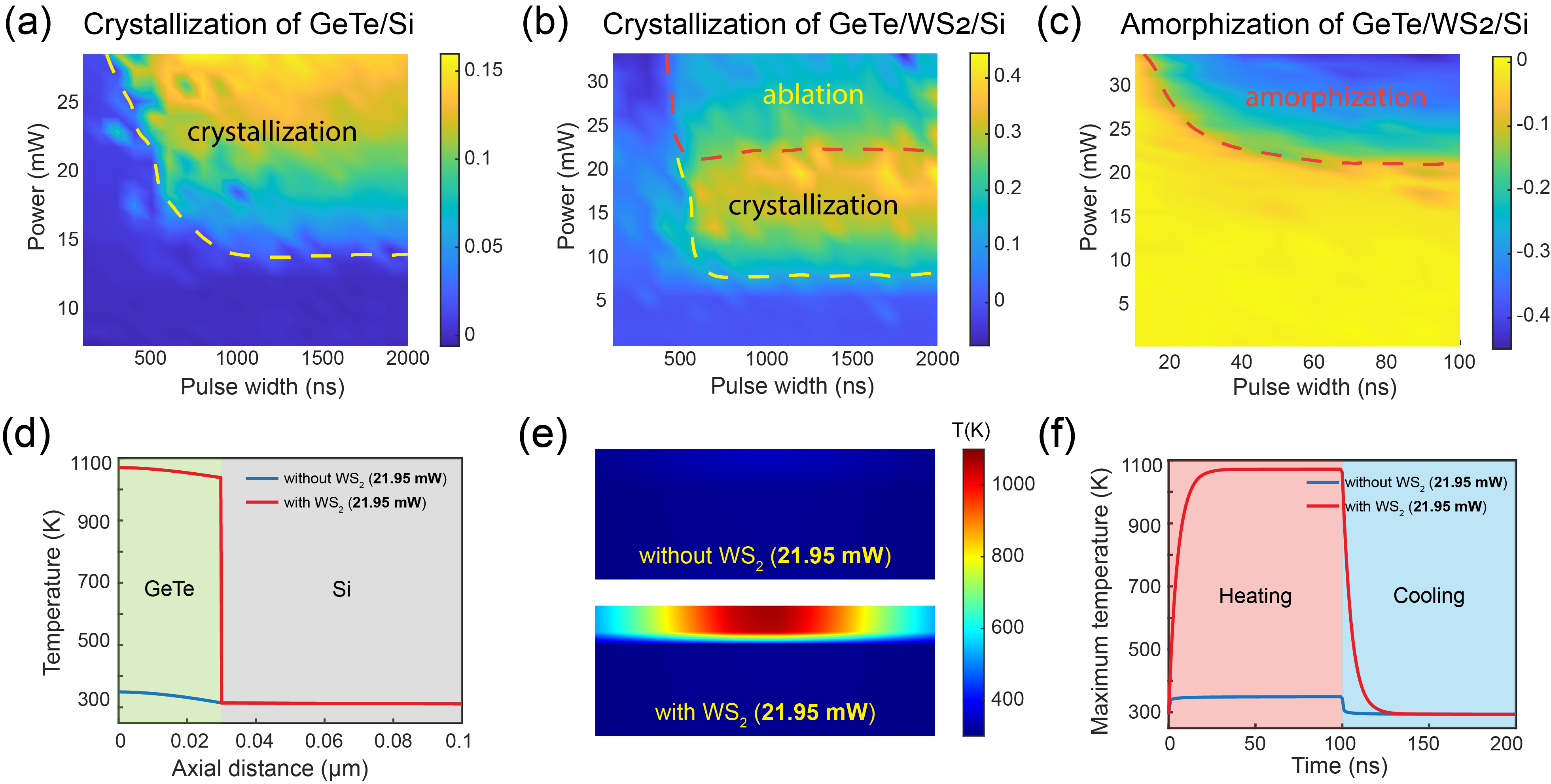}
\caption{\ch{WS2} effect on laser switching and amorphization. 
Power-time-reflectivity measurements for (a) crystallization of GeTe on Si, (b) crystallization of GeTe on \ch{WS2} on silicon, and (c) re-amorphization of GeTe on \ch{WS2} on silicon.
Simulated temperature of GeTe on \ch{WS2} on silicon during amorphization. 
(d) Axial temperature distribution in GeTe films on \ch{Si}  without and with a \ch{WS2} 2D layer using same laser pulse  power.
(e) Cross-sectional temperature distribution of GeTe on Si and GeTe on \ch{WS2} on silicon with the same laser  pulse power.
(f) Maximum temperature on the surface of GeTe on silicon and GeTe on \ch{WS2} on silicon after a 100 ns laser pulse with the same power.
}
\label{fig_5}
\end{figure}

In order to establish the generality of this thermal barrier property amongst 2D materials and distinguish the improvement in crystallization, we grew GeTe on top of another 2D material, \ch{WS2}. 
Moreover, the samples are grown on silicon, which is more relevant to silicon photonics, rather than the silica-on-silicon substrate.
\ch{WS2} has a similar structure and properties as \ch{MoS2} with weak vdWs interfaces. 
Si is 10$\times$ more thermally conductive ($\sim$140~W$\cdot$m$^{-1}\cdot$K$^{-1}$) than \ch{SiO2}, which facilitates faster heat dissipation.
In our previous measurement, the crystallization temperature of GeTe on \ch{WS2} was measured as 201$^{\circ}$C ($T_c$), as seen in Figure~\ref{fig_1}(c), which means that the \ch{WS2} monolayer also had no chemical reaction with GeTe.
Moreover, \ch{WS2} substantially decreased the switching energy of GeTe on the silica-on-silicon substrate(Figure~\ref{fig_S2}).
We should expect the influence of \ch{WS2} on the switching power to be even more dramatic for Si substrates because its thermal conductivity is an order of magnitude greater than that of \ch{SiO2}.
We measured the laser switching power and time for GeTe thin films with and without the \ch{WS2} interfacial layer on silicon.
Again, both samples came from the same GeTe sputtering batch. 
In our laser write-mark matrix switching experiment, the crystallization power is 7.45 mW for GeTe grown on \ch{WS2} while it is 13.45 mW for GeTe deposited directly on a Si substrate, as shown in Figure~\ref{fig_5}(a) and (b).
This is in agreement with simulations where we see more than 40\% reduction in the switching power for the modeled crystallization process, as shown in Figure~\ref{fig_S4}.
This also conforms with the switching efficiency improvement seen in the GeTe/\ch{MoS2} sample during amorphization.
We also studied the amorphization processes where the annealed crystalline GeTe layer with a 0.65~nm-thick interfacial \ch{WS2} layer became less reflective after being exposing to an amorphizing 21.95 mW, 50 ns laser pulse. 
Figure~\ref{fig_5}(c) shows the corresponding laser pulse power-time-reflectivity plots with the \ch{WS2} layer. 
In contrast, we found that amorphization of GeTe directly on silicon was not possible, even at our system's power limit of 33.42 mW. 
Thus, a sufficiently thick oxide layer or a 0.65~nm TMDC layer are required to limit heat transfer to the substrate.
Indeed, the \ch{WS2} layer was necessary to amorphize the GeTe layer on Si, although the laser pulse power is higher than that required for GeTe on \ch{WS2} on the silica-on-silicon substrate(12.42~mW, 80ns).
Hence, the \ch{WS2} layer is extremely effective at reducing the RESET energy on silicon substrates.
As shown by our measurements and simulations, this improvement results from the TBR at the \ch{WS2} interface layer.
The weak vdWs interaction restricts the heat generated by laser pulses from dissipating in the out-of-plane direction.
Moreover, the laser spot size controls the in-plane heat loss in a facile way.
Surprisingly, the effect of \ch{WS2} on the heat transport between the Si and the GeTe layer is so pronounced that it was even possible to ablate the GeTe layer on \ch{WS2} on Si.
This is remarkable considering that GeTe deposited directly on Si cannot even be heated to induce amorphization.
The thermal simulations of the GeTe-\ch{WS2}-Si stacks provide some insight into this substantial difference.
Figure~\ref{fig_5}(d)-(f) shows that 21.95 mW laser heating pulses cause the GeTe on silicon to marginally heat because the heat rapidly dissipates into the Si substrate. 
Indeed, after approximately 10 ns the temperature of the GeTe saturates at 349~K, which is a negligible temperature rise.  
However, adding the subnanometer-thick \ch{WS2} monolayer confines the heat within the GeTe layer and the temperature saturates at 1070~K in 30~ns.
These results explain the reason why GeTe on Si could not be laser amorphized in our laser static tester system, but could be readily amorphized when the subnanometer thick \ch{WS2} layer was inserted between the GeTe and the silicon.
This result is especially relevant to Si photonics, where the PCM is usually placed in direct contact with the Si waveguide. 

\begin{figure}
\centering
\includegraphics[width=\textwidth]{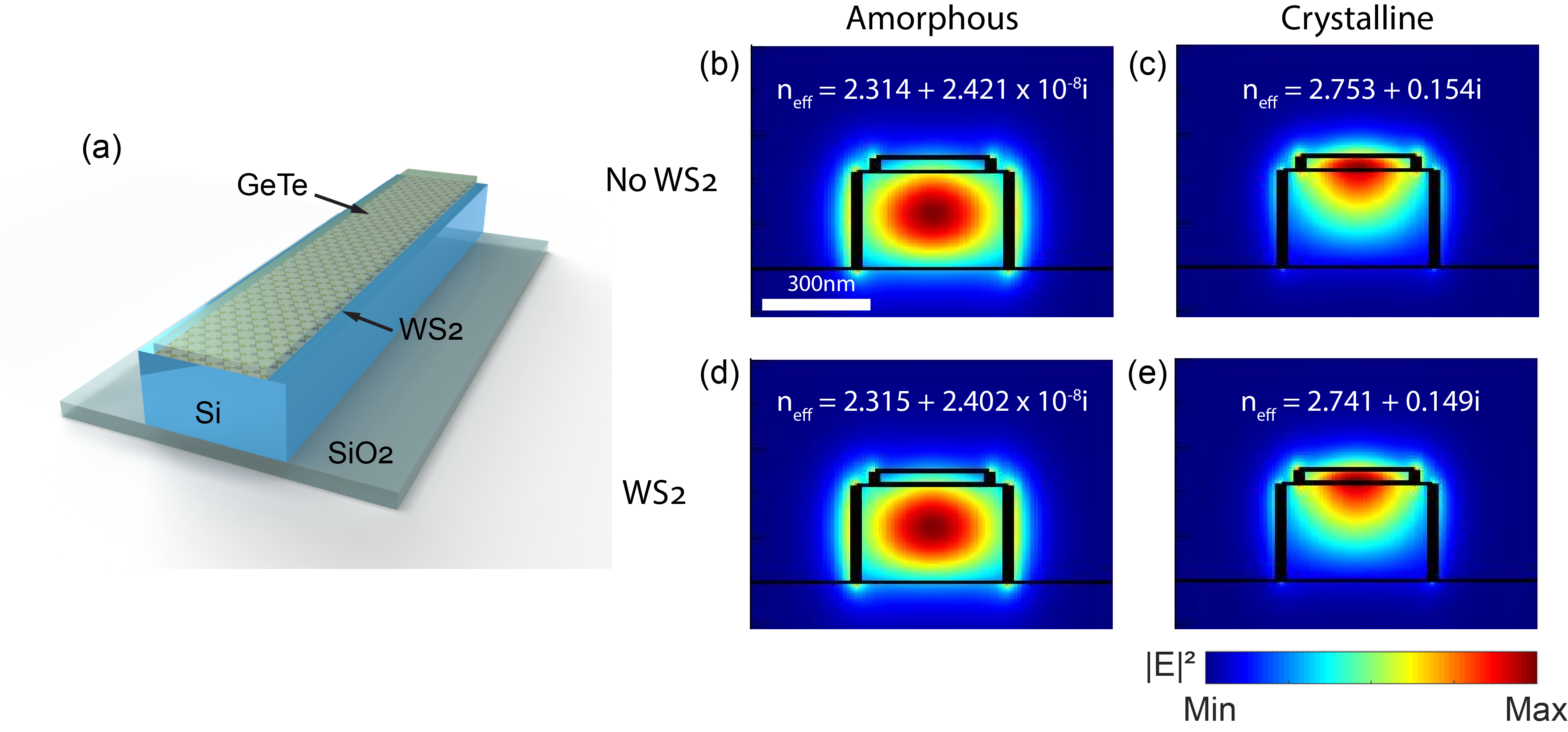}
\caption{
(a)Schematic of the GeTe-tuned Si waveguide  model with a\ch{WS2} 2D TBR layer. 
The corresponding mode patterns and effective refractive index, n$_{\textmd{eff}}$, values for amorphous and crystalline GeTe without and with \ch{WS2} are shown in (b)-(e).
}
\label{fig_6}
\end{figure}

These results indicate that incorporating subnanometer thick TMDC layers into PCM-based reconfigurable Photonic Integrated Circuit (PIC) devices will allow efficient PCM switching. 
However, ideally the TMDC material should have negligible interaction with the optical mode propagating in the waveguide. 
To demonstrate the compatibility of sub-nanometer thick TMDCs with photonic devices, we compare the changes in optical mode confinement of photonic waveguides with and without the TMDC layer using Finite Difference Eigenmode (FDE) calculations. 
Figure~\ref{fig_6}(a) shows the waveguide simulation model with the \ch{WS2} layer. 
The resulting mode pattern and effective refractive index, n$_{\textmd{eff}}$, of the waveguides at the 1550 nm wavelength, which is in the telecommunication c-band, are shown in Figure~\ref{fig_6}(b)-(e). 
We observe that the thin \ch{WS2} layer negligibly changes the real part of the effective refractive index by less than 0.3\% for both amorphous and crystalline GeTe-tuned waveguides. 
The absolute effective refractive index values are also shown in Figure~\ref{fig_6}(b)-(e).
Moreover, there are no discernible changes in the mode patterns.
The non-discernible change in mode pattern is partly due to \ch{WS2} having a close refractive index value to Si and being non-absorbing in the infrared due to its large bandgap(Figure~\ref{fig_S5})\citep{EDWARDS1985,liu2020temperature}. 
Therefore, the \ch{WS2} layer can be incorporated into PCM-based PIC devices with minimal optical effect yet produce a dramatic reduction in the PCM switching energy, which is a highly desirable trait.
Since the \ch{WS2} on silica-on-silicon substrates showed a similar amorphization performance to that of \ch{MoS2} on silica-on-silicon, we also expect \ch{MoS2} to show a similar improvement as \ch{WS2} if it is placed directly on the Si waveguide.
Similarly, a \ch{MoS2} interfcial 2D layer causes a negligibe change in the effective refractive index and concomitant modes of the GeTe-tuned PCM waveguide (Figure~\ref{fig_S6}).

\section{Conclusions}
In conclusion, we have demonstrated a radical reduction in the switching energy of GeTe on different substrates by employing interfacial subnanometer thick TMDC 2D crystal layers. 
We expect these performance enhancements to be broadly applicable to programmable photonics, especially programmable plasmonic metamaterials and Si photonics, where the PCM is often placed in direct contact with materials of high thermal conductivity.
The enhancement in switching energy efficiency is due to the 2D material vdWs bonds confining heat in the PCM layer. 
We demonstrated that the PCMs integrated 2D layer consumed less energy in both optical crystallization and amorphization operations. 
There is an over 40\% reduction in power when the \ch{MoS2} layer is interfaced between the PCM and a silica-on-silicon substrate. 
The improvement when \ch{WS2} is placed on Si is even more pronounced but we were not able to quantify the enhancement because without the TMDC 2D layer, the PCM could not even be switched due to the power requirement being too high. 
However, simulations show that the equilibrium temperature for 21.95 mW laser pulses is increased by more than 700 K when a \ch{WS2} layer is included between the GeTe layer and the Si.
We found that in PCM-programmed Si waveguide simulations, these 2D TMDC layers have a negligible effect on the mode pattern and the waveguide effective refractive index.
These results show that 2D TMDC layers should be included when designing efficient PCM-programmable devices, such as photonic memories, all-optical neural networks, and plasmonic metasurfaces.

\begin{acknowledgement}
The SUTD research was funded by a Singapore MoE Project "Electric-field induced transitions in chalcogenide monolayers and superlattices", grant MoE 2017-T2-1-161 and an A*STAR AME project: "Nanospatial Light Modulators (NSLM)", A18A7b0058. 
The 2D materials work is supported by the UK's Engineering and Physical Sciences Research Council through the Future Photonics Manufacturing Hub (EPSRC EP/N00762X/1), the Chalcogenide Photonic Technologies (EPSRC EP/M008487/1) and ChAMP--Chalcogenide Advanced Manufacturing Partnership (EPSRC EP/G060363/1).
Ms Jing Ning is grateful for her MoE PhD scholarship.

\end{acknowledgement}

\begin{suppinfo}


\end{suppinfo}

\bibliography{2D_refer}
\makeatletter\@input{si.tex}\makeatother
\end{document}


\begin{figure}
\centering
\includegraphics[width=\textwidth]{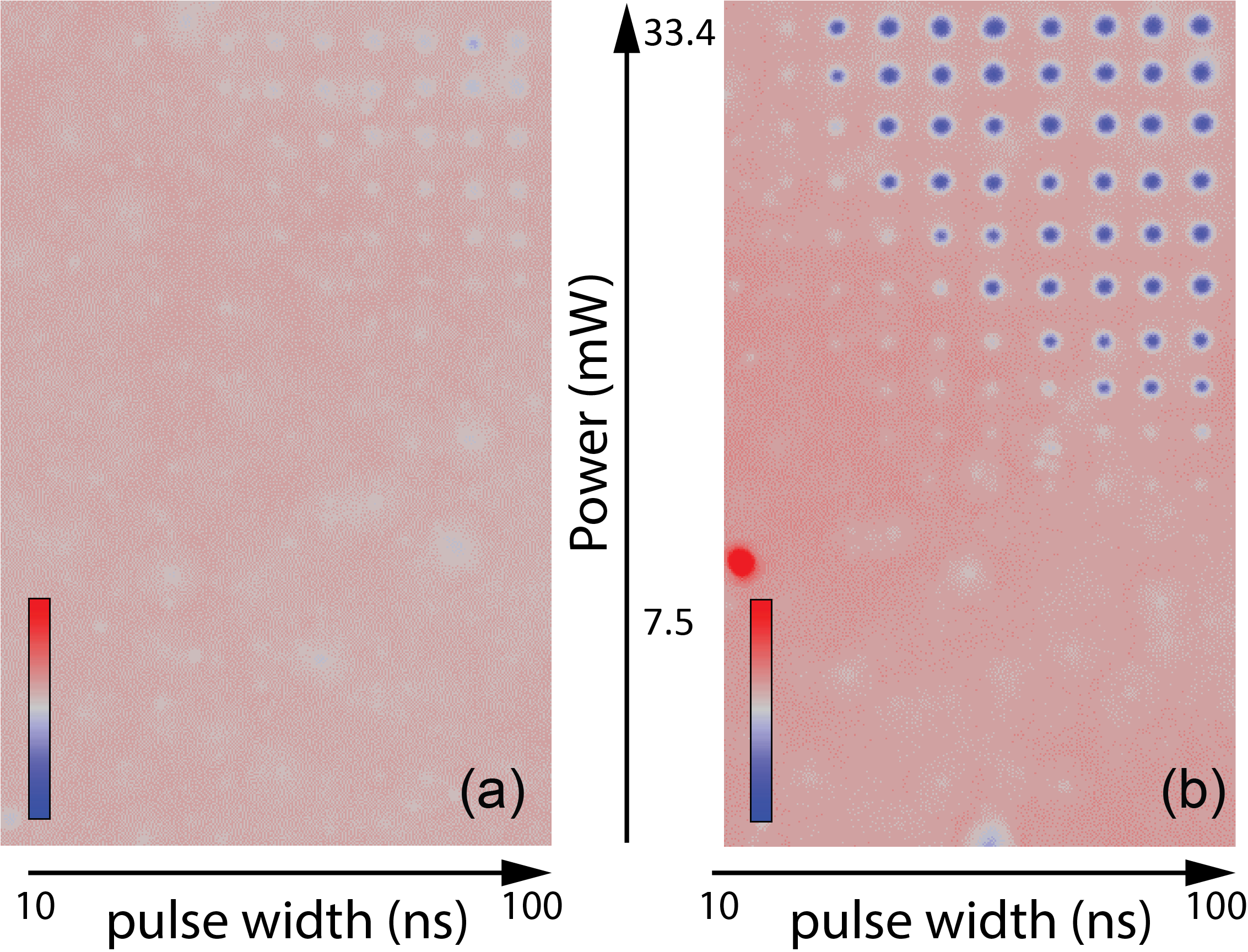}
\caption{ Optical micrographs of amorphization matrix on annealed crystalline GeTe (a) and GeTe with \ch{MoS2} (b) samples. 
The red color represent the high reflective state, and the blue color represents the low reflective state. 
}
\label{fig_S1}
\end{figure}

\begin{figure}
\centering
\includegraphics[width=\textwidth]{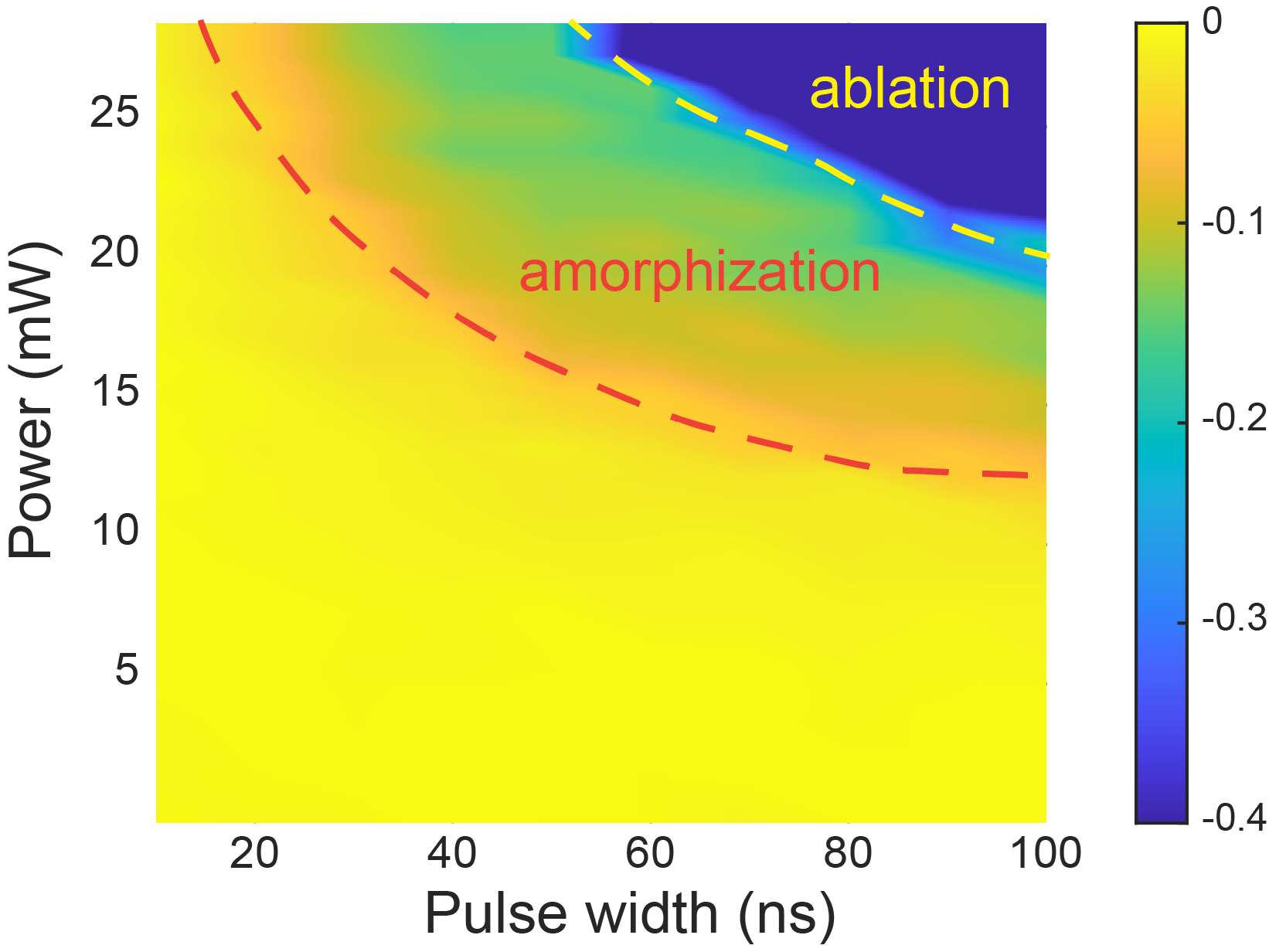}
\caption{ 
Power-time-reflectivity measurement of GeTe with \ch{WS2} on \ch{SiO2/Si} substrate in amorphization.
}
\label{fig_S2}
\end{figure}

\begin{figure}
\centering
\includegraphics[width=\textwidth]{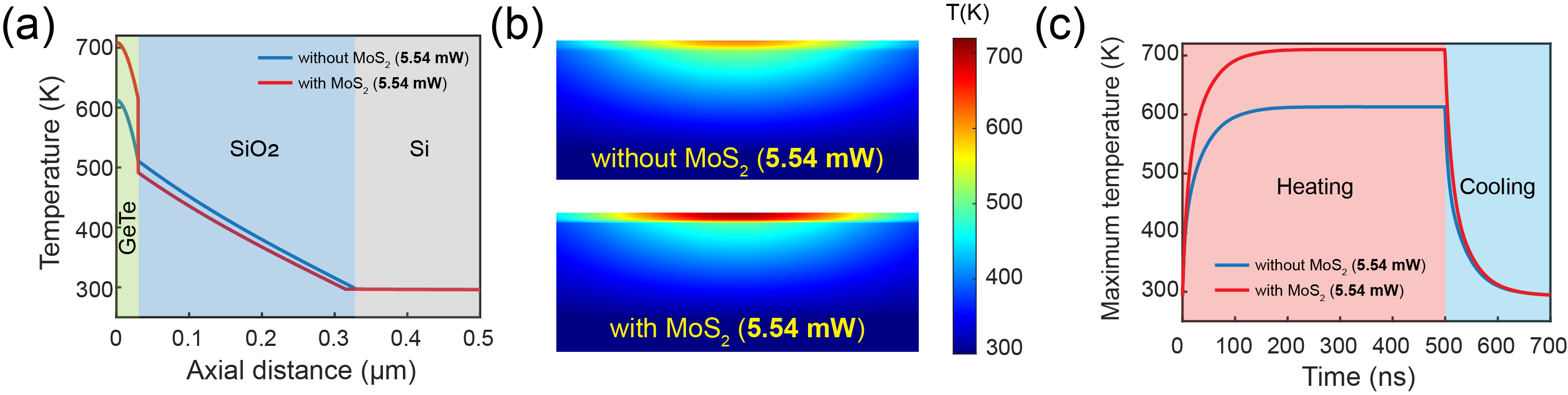}
\caption{Heat transport simulation of crystallization and amorphization behavior. 
(a) Axial temperature distribution in GeTe films on \ch{SiO2/Si} substrate without and with \ch{MoS2} monolayer after a 500 ns-pulse with different input power.
(b) Temperature distribution in cross-section of GeTe film and GeTe on \ch{MoS2} after a 500 ns-pulse with different power pulse.
(c) Maximum temperature on the surface with different power.
}
\label{fig_S3}
\end{figure}

\begin{figure}
\centering
\includegraphics[width=\textwidth]{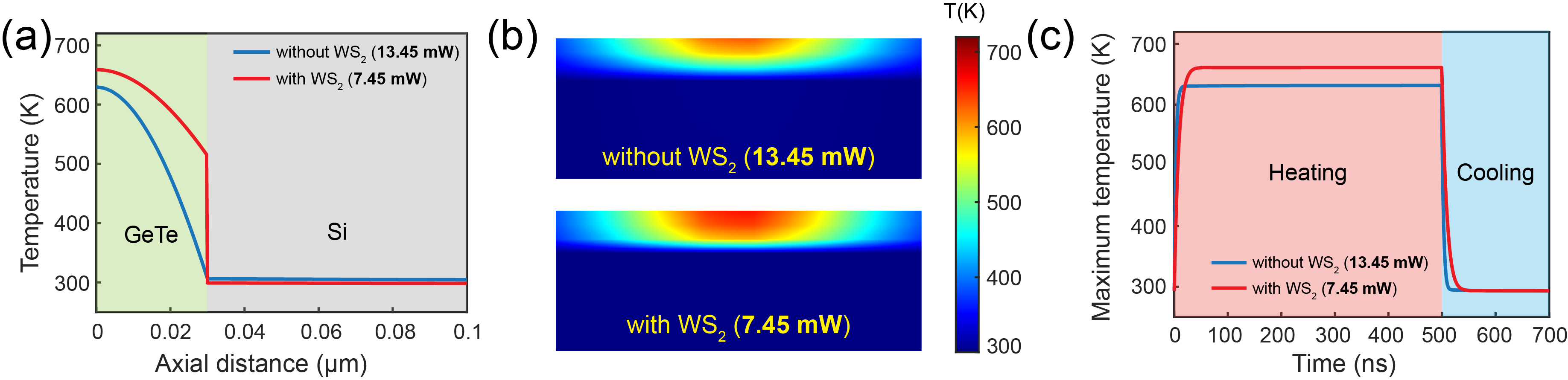}
\caption{Heat transport simulation of GeTe on \ch{WS2} layer in crystallization.
(a) Axial temperature distribution in GeTe films on \ch{Si} substrate without and with \ch{WS2} monolayer after a 500 ns-pulse with different input power.
(b) Temperature distribution in cross-section of GeTe film and GeTe on \ch{WS2} after a 500 ns-pulse with different power pulse.
(c) Maximum temperature on the surface of GeTe and GeTe on \ch{WS2} samples after a 500 ns-pulse with different power.
}
\label{fig_S4}
\end{figure}

\begin{figure}
\centering
\includegraphics[width=\textwidth]{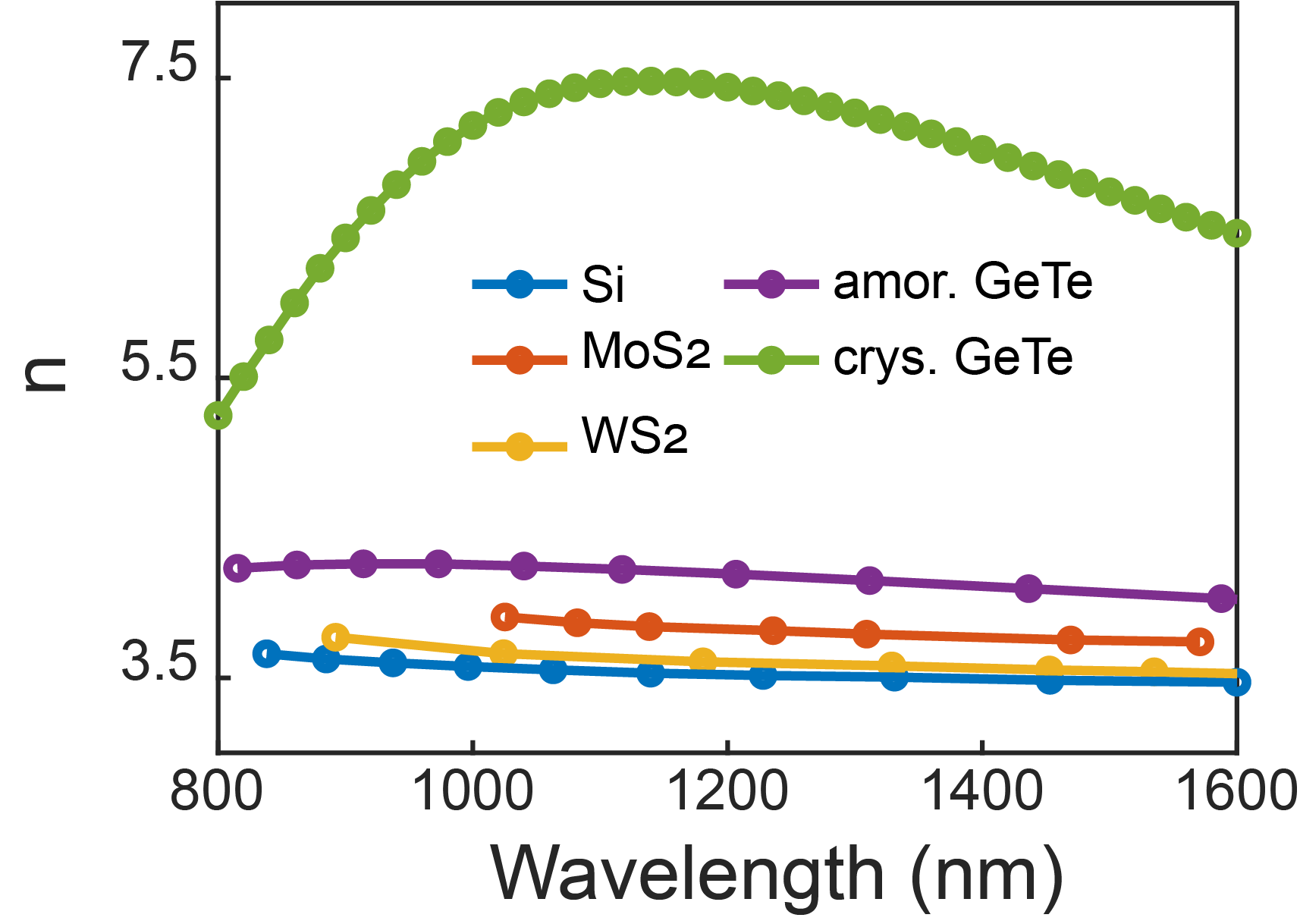}
\caption{Refractive index of the materials used in the waveguide simulation model. 
}
\label{fig_S5}
\end{figure}

\begin{figure}
\centering
\includegraphics[width=\textwidth]{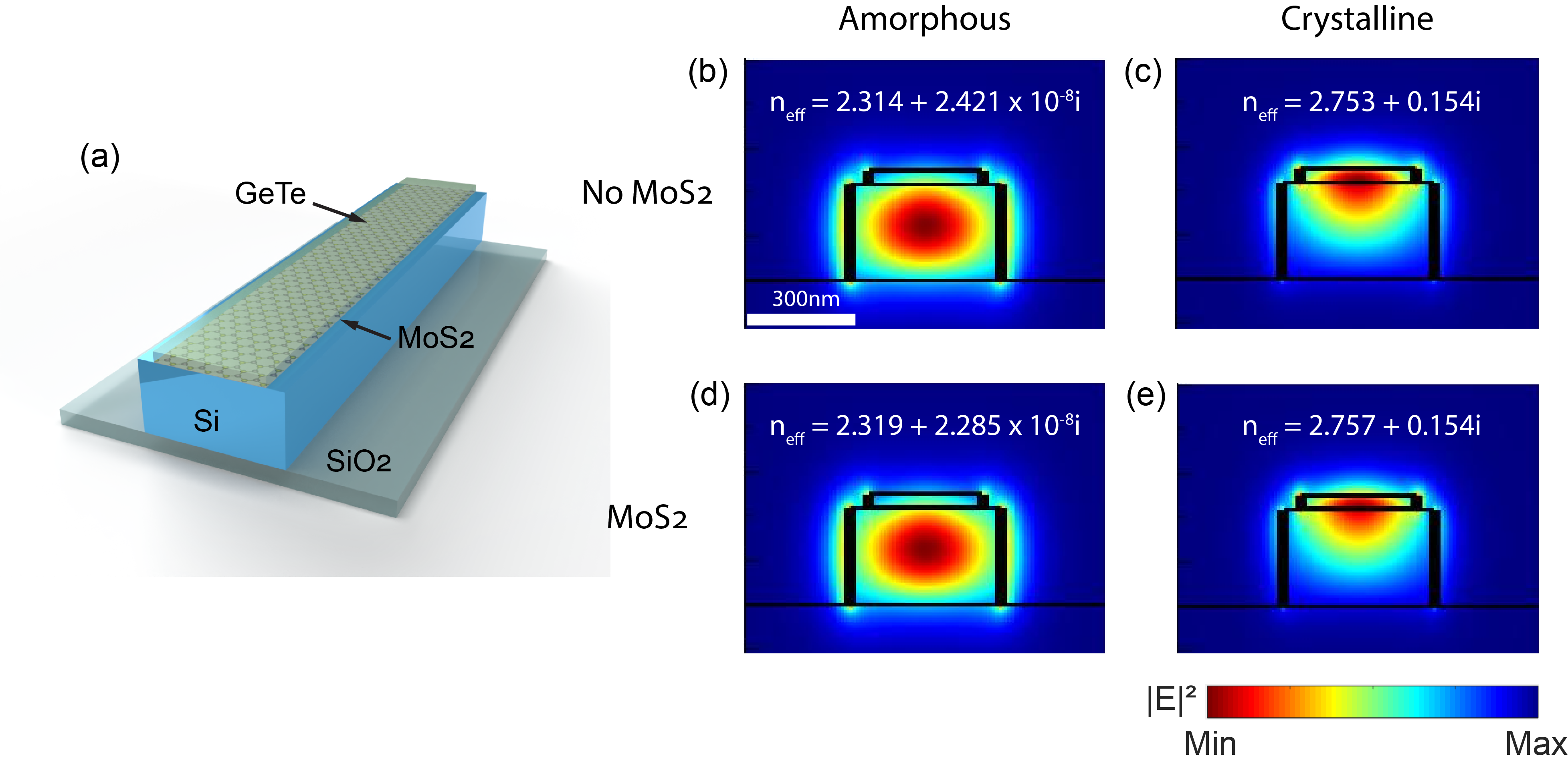}
\caption{
(a)Schematic of GeTe-tuned Si waveguide simulation model with \ch{MoS2} layer. 
The corresponding mode patterns and effective refractive index, n$_{\textmd{eff}}$, values for amorphous and crystalline GeTe without or with \ch{MoS2} are shown in (b)-(e).
}
\label{fig_S6}
\end{figure}